\documentclass[useAMS,usenatbib,usegraphicx]{mn2e}
\usepackage{pslatex,color}
\usepackage{amsmath,amsfonts,amssymb,amsxtra,eufrak,url}

\def\idm#1{{\mbox{\scriptsize #1}}}
\newcommand{\au}{\mbox{au}}
\newcommand{\msun}{\mbox{M}_{\sun}}

\usepackage{color}
\definecolor{myred}{rgb}{0.7,0.1,0.1}
\definecolor{myblue}{rgb}{0.2,0.0,0.7}
\definecolor{mybrown}{rgb}{0.5,0.2,0.0}

\newcommand\corr[1]{{#1}}

\newcommand\hide[1]{}
\title[A linear distribution of orbits in compact planetary systems?]
{A linear distribution of orbits in compact planetary systems?}
\author[Migaszewski et al.]
{Cezary Migaszewski$^{1}$,
Krzysztof Go\'zdziewski$^{1}$ 
\& Mariusz S{\l}onina$^{1}$\\
$^{1}$ Centre for Astronomy, Faculty of Physics, Astronomy and Informatics, Nicolaus Copernicus University, Grudziadzka 5, 87-100 Torun, Poland}
\begin{document}
%
\date{Accepted 2013 August 01. Received 2013 July 12; in original form 2013 April 30}
\pagerange{\pageref{firstpage}--\pageref{lastpage}} \pubyear{2013}
\maketitle
\label{firstpage}
%
%
\begin{abstract}
We report a linear ordering of orbits in a sample of multiple extrasolar
planetary systems with super-Earth planets.  We selected $20$ 
cases, mostly discovered by the Kepler mission, hosting at least
four planets within $\sim 0.5\,\au$. The semi-major axis $a_n$
of an $n$-th planet in each system of this sample obeys $a(n) = a_1 +
(n-1)\,\Delta\,a$, where $a_1$ is the semi-major axis of the innermost orbit
and $\Delta\,a$ is a spacing between subsequent planets,
which are specific for a particular system. 
For instance, the Kepler-33
system hosting five super-Earth planets exhibits the relative deviations
between the observed and linearly predicted semi-major axes of only a few
percent. At least  half of systems in the sample fulfill the linear law with a
similar accuracy. We explain the linear distribution of semi-major
axes as a natural implication of multiple chains of mean motion
resonances between subsequent planets, which emerge
due to planet--disk interactions and convergent migration at
early stages of their evolution.
\end{abstract}
%
%
\section{Introduction}
%
The Kepler photometric mission \citep{Borucki2010} brought many discoveries
of multiple low-mass planetary systems. There are several known systems with
four or more super-Earths and/or Neptune/Uranus mass planets. In particular there
are the Kepler-11 system with six planets \citep{Lissauer2011}, five-planet
systems Kepler-33 \citep{Lissauer2012}, Kepler-20 \citep{Gautier2012} and
Kepler-32 \citep{Fabrycky2012}. There are also a few systems with five
candidate planets: KOI-435 \citep{Ofir2012}, KOI-500, KOI-505
\citep{Borucki2011} and several more four-planet systems. Configurations of
this \corr{type} were first discovered with Doppler spectroscopy, e.g., 
Gliese~876 \citep{Rivera2010},
Gliese~581 \citep{Forveille2011}, 
HD~10180 \citep{Lovis2011}
and HD~40307 \citep{Tuomi2013}. 
In the Gliese~876 \corr{system}, however, two of the companions are 
jovian planets, similarly to the Kepler-94 system.
All studied systems, with a few  discussed
furthermore, host at least four planets with orbital semi-major axes  $\lesssim 0.5\,\au$.

These discoveries raise a question on mechanisms 
leading to such compact ordering of the planetary systems, and simultaneously 
providing their long-term
stability.  Our recent study of the Kepler-11 system \citep{Migaszewski2012}
revealed that this configuration of six super-Earths is chaotic,
and its marginal, long-term dynamical (Lagrangian) stability is most likely possible 
due to particular
multiple mean-motion resonances between the planets.  
In the sample quoted above, we may pick up
multiple configurations even more compact, and bounded 
to the distance as small as $0.08\,\au$,
e.g., Kepler's KOI-500 with five planets.  
The orbital architecture of planetary systems of this
class recalls the hypothesis of the Packed Planetary Systems
\cite[PPS,][]{Barnes2004}, though originally formulated for configurations with 
jovian companions. 
In the jovian mass range, the orbital stability of multiple systems
is statistically preserved, if planets in subsequent pairs
with semi-major axes $a_1, a_2$ and masses $m_1, m_2$ are separated by more 
than $K \sim 4,5$ mutual Hill radii $R_{H,M}$, where 
$R_{H,M}=(1/2) (a_1+a_2) [(m_1+m_2)/(3 m_*)]^{(1/3)}$ and $m_*$ is the mass
of the parent star
\citep{Chatterjee2008}. However, $R_{H,M}$ in the above systems
is of the order of $10^{-3}$~au, hence their typical separation 
is \corr{at least} one order of magnitude larger, $K\sim 10$,
\corr{for the outermost pairs of planets, while for the innermost
planets $K\sim 30$--$50$}. \corr{A study of systems with 1 Earth-mass planets orbiting a Sun-like star conclude that the stability
is maintained for $K$ roughly larger than 10-13 \citep{Smith2009}. This seems in accord with
the analysis of the Kepler-11 \citep{Migaszewski2012} and similar systems
 which reveals, that likely
they evolved into a particular architectures helping to maintain the stability}. Indeed, due to
small eccentricities, these systems unlikely suffered
planet-planet scattering, \corr{often quoted in the literature to
explain the observed eccentricity distribution 
in the sample of multiple extrasolar systems \cite[e.g.][]{Raymond2009}}.  
In the light of the PPS hypothesis, the multiple, compact systems with
super-Earths will be classified as packed multiple-planet systems, from
hereafter.
In this Letter, we report a detection of a 
linear ordering of the planets with their number (index) and argue that \corr{such particular architecture might stem} from the planetary migration.

%
\section{A prototype case: the Kepler-33 system}
%
The Kepler-33 system hosts five planets. For the parent star mass $m_0 = (1.29 \pm
0.12)\,\msun$ (see caption to Table~\ref{tab:tab1} for references to all discussed systems) and the reported orbital periods, we computed the semi-major
axes of the planets, $a_n$, where $n=2,3,4,5,6$. The plot of
$a_n$ against $n$ (the top left-hand panel of Fig.~\ref{fig:chosen_an})
reveals a clear linear correlation $a(n) = 0.024 +
0.047\,(n-1)$ (shown as a green line). 
We start the sequence of indexes from $2$ rather than from $1$ to have $a_1
\in [0, \Delta\,a)$.  We want this condition to be fulfilled in all studied
examples.  The uncertainties of the best-fit parameters $a_1, \Delta\,a$
(accompanied by other quantities introduced below) are given in the first
row of Table~\ref{tab:tab1}.
All of $a_n$ are very close to the line on the
$(n, a_n)$-graph. In the sample of multiple Kepler systems, we found a
few other systems exhibiting a similar dependence of the semi-major axes on
the planet index. To express deviations between observed and 
predicted semi-major axis (O-C) of a planet in a given system, we introduce
$
\Delta_n \equiv [a_n - a(n)]/a_n
$ and
$
\bar{\Delta}_n \equiv [a_n - a(n)]/\Delta\,a,
$
which are the (O-C) scaled by $a_n$ and $\Delta\,a$, respectively. 
The top-left panel of Fig.~1 is labeled by
$\Delta_n$ and $\bar{\Delta}_n$ expressed in percents, close to each red
filled-circle marking a particular $a_n$. Values of $\Delta_n$ are given
above the linear graph, and $\bar{\Delta}_n$ below the graph.
To measure the ''goodness of fit'' of the linear model for a whole 
$N$-planet system, we define 
$
\delta \equiv \big[\frac{1}{N} \sum_{i=1}^{N} \bar{\Delta}_{n(i)}^2\big]^{1/2} \times 100\,\%,
$
where $n(i)$ is an index given to $i$-th
planet. Therefore, $\delta$ is equivalent to the common rms
scaled by the spacing parameter $\Delta a$. 
When the indexes $n(i)$ for subsequent planets of the Kepler-33 are $2, 3, 4, 5, 6$, the resulting 
$\delta \approx 6.1\,\%$. We did not find any better 
linear fit parameters and planets numbering. However, in other cases,
as shown below,
non-unique solutions may appear for different $\Delta\,a$.

To find the best-fit combination of the $a_1,\Delta a$ and a sequence of indexes $\left\{ n(i)\right\}_{i=1}^{i=N}$, for each studied system, we perform a simple optimization. We fix a point in the $(a_1,\Delta\,a)$--plane, where $a_1 \in [0, \Delta\,a)$ and look for a set of indexes $n(i)$ providing $\min \bar{\Delta}_{n(i)}$.
The results for the Kepler-33 system are illustrated in the top left-hand panel of Fig.~\ref{fig:chosen_delta2} in the form of one-dimensional scan over $\Delta a$. 

\begin{table*}
\caption{The results of analysis of a sample of $21$ systems. References: $1$: \protect\cite{Lissauer2012}, $2$: \protect\cite{Ofir2012}, $3$: \protect\cite{Borucki2011}, $4$: \protect\cite{Fabrycky2012}, $5$: \protect\cite{Tuomi2013}, $6$: \protect\cite{Hirano2012}, $7$: \protect\cite{Weiss2013}, $8$: \protect\cite{Lissauer2011}, $9$: \protect\cite{Forveille2011}, $10$: \protect\cite{Gautier2012}, 
$11$: \protect\cite{Rivera2010}, 
$12$: \protect\cite{Lovis2011}.
}
\label{tab:tab1}
\begin{tabular}{l c c c c c c c c c}
\hline
\hline
star & $m_0 [\msun]$ & $N$ & ref. & $\Delta\,a\,[\au]$ & $a_1\,[\au]$ & $\delta [\%]$ & $f_{2/3} [\%]$ & $f_1 [\%]$ & sequence \\
\hline
Kepler-33 & $1.29 \pm 0.12$ & $5$ & $1$ & $0.0466 \pm 0.0012$ & $0.024 \pm 0.004$ & $6.1$ & $5.8$ & $1.8$ & $2-3-4-5-6$ \\
KOI-435 & $0.9$ & $5$ & $2$ & $0.0507 \pm 0.0012$ & $0.0419 \pm 0.0033$ & $6.9$ & $8.0$ & $2.5$ & $1-2-3-4-6$ \\
KOI-1955 & $1.0^*$ & $4$ & $2$ & $0.0497 \pm 0.0015$ & $0.025 \pm 0.004$ & $6.0$ & $14.4$ & $6.6$ & $1-3-4-5$\\
KOI-719 & $0.68$ & $4$ & $3$ & $0.0291 \pm 0.0004$ & $0.0155 \pm 0.0017$ & $4.3$ & $8.0$ & $3.6$ & $2-3-6-8$\\
KOI-408 & $1.05$ & $4$ & $3$ & $0.03006 \pm 0.00035$ & $0.0157 \pm 0.0013$ & $3.0$ & $3.8$ & $1.7$ & $2-3-4-7$\\
KOI-671 & $0.96$ & $4$ & $3$ & $0.02428 \pm 0.00082$ & $0.001 \pm 0.003$ & $5.7$ & $11.3$ & $5.2$ & $3-4-5-6$\\
Kepler-32 & $0.58 \pm 0.05$ & $5$ & $4$ & $0.01952 \pm 0.00036$ & $0.0132 \pm 0.0012$ & $6.5$ & $6.7$ & $2.1$ & $1-2-3-4-7$\\
KOI-500 & $0.66$ & $5$ & $3$ & $0.0146 \pm 0.0006$ & $0.004 \pm 0.002$ & $10.2$ & $23.2$ & $7.9$ & $2-3-4-5-6$\\
HD 40307 & $0.77 \pm 0.05$ & $6$ & $5$ & $0.02795 \pm 0.00016$ & $0.0229 \pm 0.0015$ & $7.2$ & $4.1$ & $0.9$ & $2-3-5-7-9-22$\\
KOI-730 & $1.07$ & $4$ & $3$ & $0.01400 \pm 0.00044$ & $0.0071 \pm 0.0033$ & $8.3$ & $26.8$ & $12.8$ & $6-7-9-11$\\
KOI-94 & $1.25 \pm 0.40$ & $4$ & $6, 7$ & $0.0637 \pm 0.0027$ & $0.0437 \pm 0.0062$ & $8.8$ & $29.0$ & $13.8$ & $1-2-3-5$\\
Gliese 581 & $0.31 \pm 0.02$ & $5$ & $9$ & $0.01469 \pm 0.00013$ & $0.013 \pm 0.001$ & $7.3$ & $9.7$ & $3.1$ & $2-3-5-10-15$\\
KOI-510 & $1.03$ & $4$ & $3$ & $0.0243 \pm 0.0005$ & $0.0184 \pm 0.0022$ & $7.4$ & $21.4$ & $10.1$ & $2-3-5-9$\\
KOI-505 & $1.01$ & $5$ & $3$ & $0.01181 \pm 0.00006$ & $0.008 \pm 0.001$ & $9.8$ & $20.3$ & $6.9$ & $4-6-7-10-33$\\
Kepler-31 & $1.21 \pm 0.17$ & $4$ & $4$ & $0.0521 \pm 0.0013$ & $0.047 \pm 0.006$ & $8.0$ & $24.8$ & $11.8$ & $2-3-5-8$\\
 &  &  &  & $0.0807 \pm 0.0033$ & $0.008 \pm 0.010$ & $8.3$ & $26.0$ & $12.3$ & $2-3-4-6$\\
Kepler-11 & $0.95 \pm 0.10$ & $6$ & $8$ & $0.01427 \pm 0.00011$ & $0.0077 \pm 0.0019$ & $13.4$ & $33.9$ & $8.8$ & $7-8-11-14-18-33$\\
 &  &  &  & $0.0519 \pm 0.0013$ & $0.0453 \pm 0.0051$ & $11.7$ & $22.5$ & $5.5$ & $2-2-3-4-5-9$\\
Kepler-20 & $0.912 \pm 0.035$ & $5$ & $10$ & $0.0149 \pm 0.0001$ & $0.0027 \pm 0.0011$ & $7.9$ & $11.8$ & $3.8$ & $4-5-7-10-24$\\
 &  &  &  & $0.0232 \pm 0.0003$ & $0.0211 \pm 0.0019$ & $9.6$ & $19.3$ & $6.5$ & $2-3-4-6-15$\\
KOI-623 & $1.21$ & $4$ & $3$ & $0.0353 \pm 0.0027$ & $0.024 \pm 0.008$ & $11.8$ & $47.3$ & $23.8$ & $2-3-4-5$\\
 &  &  &  & $0.0282 \pm 0.0017$ & $0.013 \pm 0.007$ & $12.2$ & $49.4$ & $24.9$ & $3-4-5-7$\\
Gliese 876 & $0.33 \pm 0.03$ & $4$ & $11$ & $0.1030 \pm 0.0063$ & $0.021 \pm 0.012$ & $9.5$ & $33.5$ & $16.1$ & $1-2-3-4$\\
 &  &  &  & $0.064 \pm 0.003$ & $0.015 \pm 0.009$ & $10.9$ & $41.6$ & $20.3$ & $1-3-4-6$\\
HD 10180 & $1.06 \pm 0.05$ & $5$ & $12$ & $0.0717 \pm 0.0003$ & $0.0597 \pm 0.0027$ & $4.8$ & $3.0$ & $1.0$ & $1-2-4-7-20$\\
 &  & $8$ &  & $0.02246 \pm 0.00006$ & $0.0202 \pm 0.0014$ & $11.3$ & $9.0$ & $1.0$ & $1-3-4-6-12-15-22-63$\\
\hline
\hline
\end{tabular}
\end{table*}

Red color is for solutions for which minimal difference between subsequent indexes 
equal to $1$. For instance, a solution of a given $\Delta\,a$ corresponding to a sequence $1-2-4-6-7$ would be plotted 
in red, because differences between indexes $n(2)=2$ and $n(1)=1$ as well as $n(5)=7$ and $n(4)=6$ equals $1$. On the other hand, $\Delta\,a$ corresponding to a sequence $1-3-5-8-11$ would be plotted in black (minimal difference between indexes equals $2$ in this case). It is obvious that when $\Delta\,a$ is much smaller than the distance between planets forming the closest pair in a system, one can obtain 
apparently very low values of $\delta$. To avoid such artificial solutions, we limit our analysis to solutions from the red part of scans presented on Fig.~\ref{fig:chosen_delta2}.

%
%
\subsection{Testing the linear ordering for known packed systems}
\begin{figure*}
\vbox{
\centerline{
\includegraphics[width=0.310\textwidth]{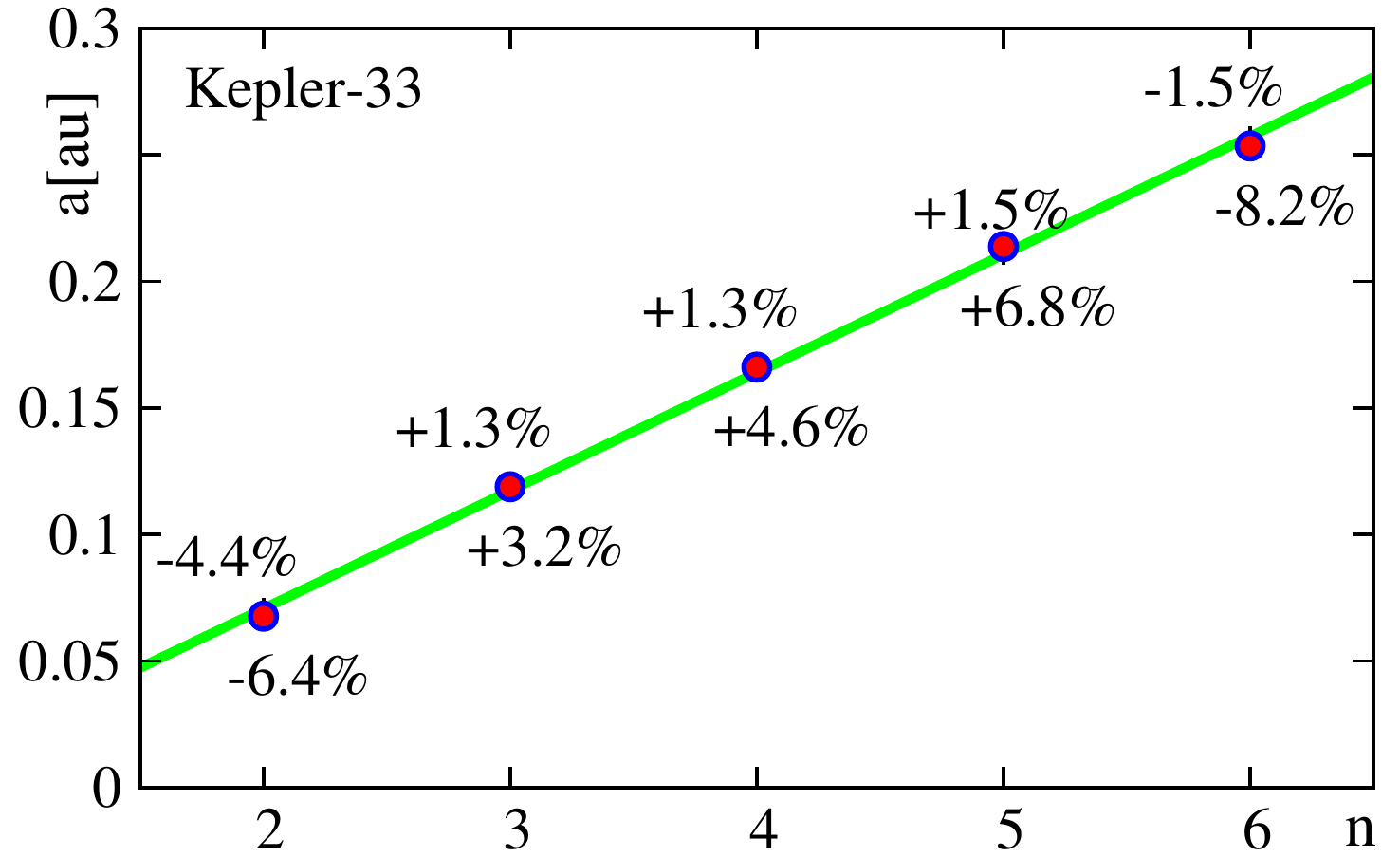}
\includegraphics[width=0.310\textwidth]{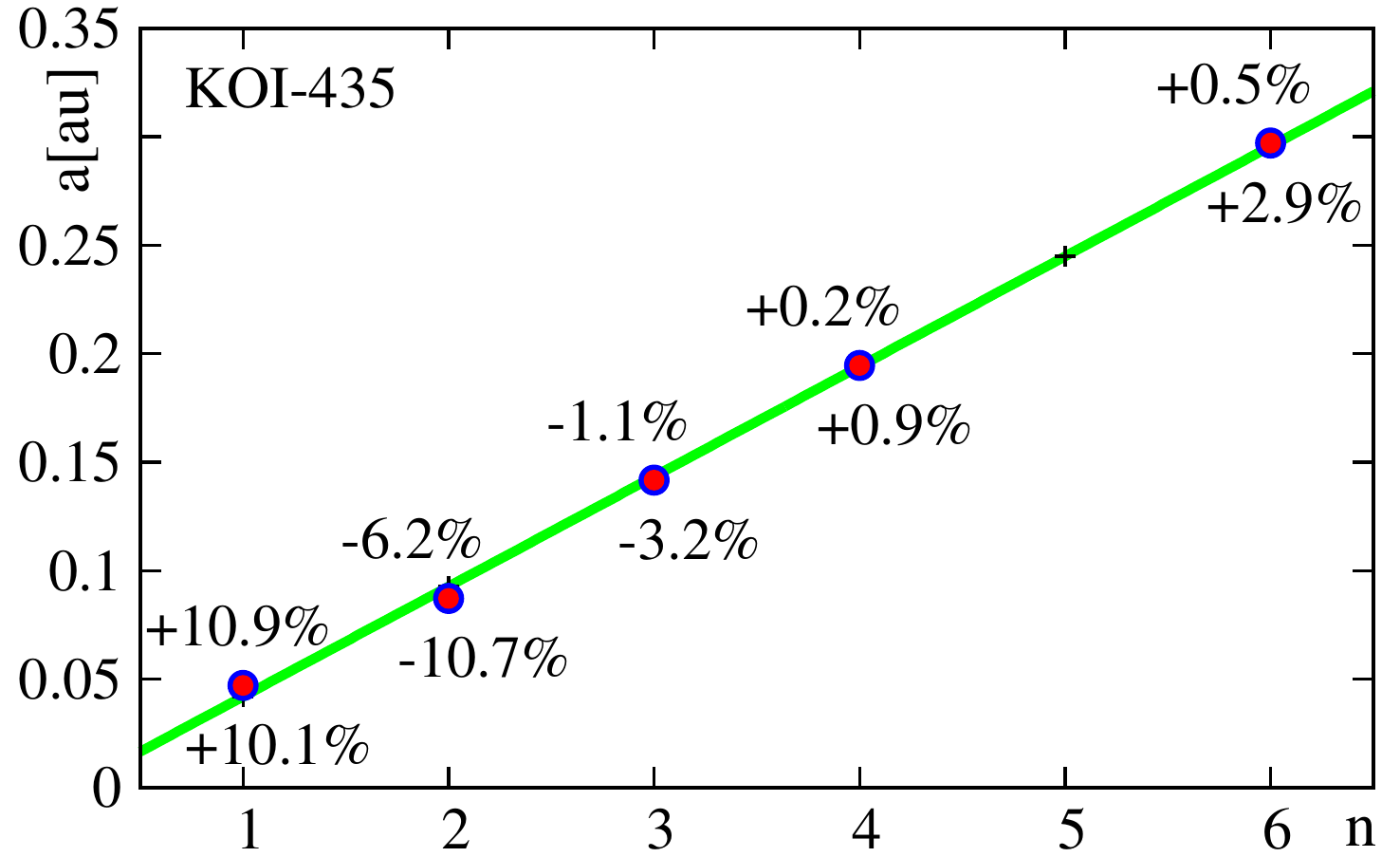}
\includegraphics[width=0.310\textwidth]{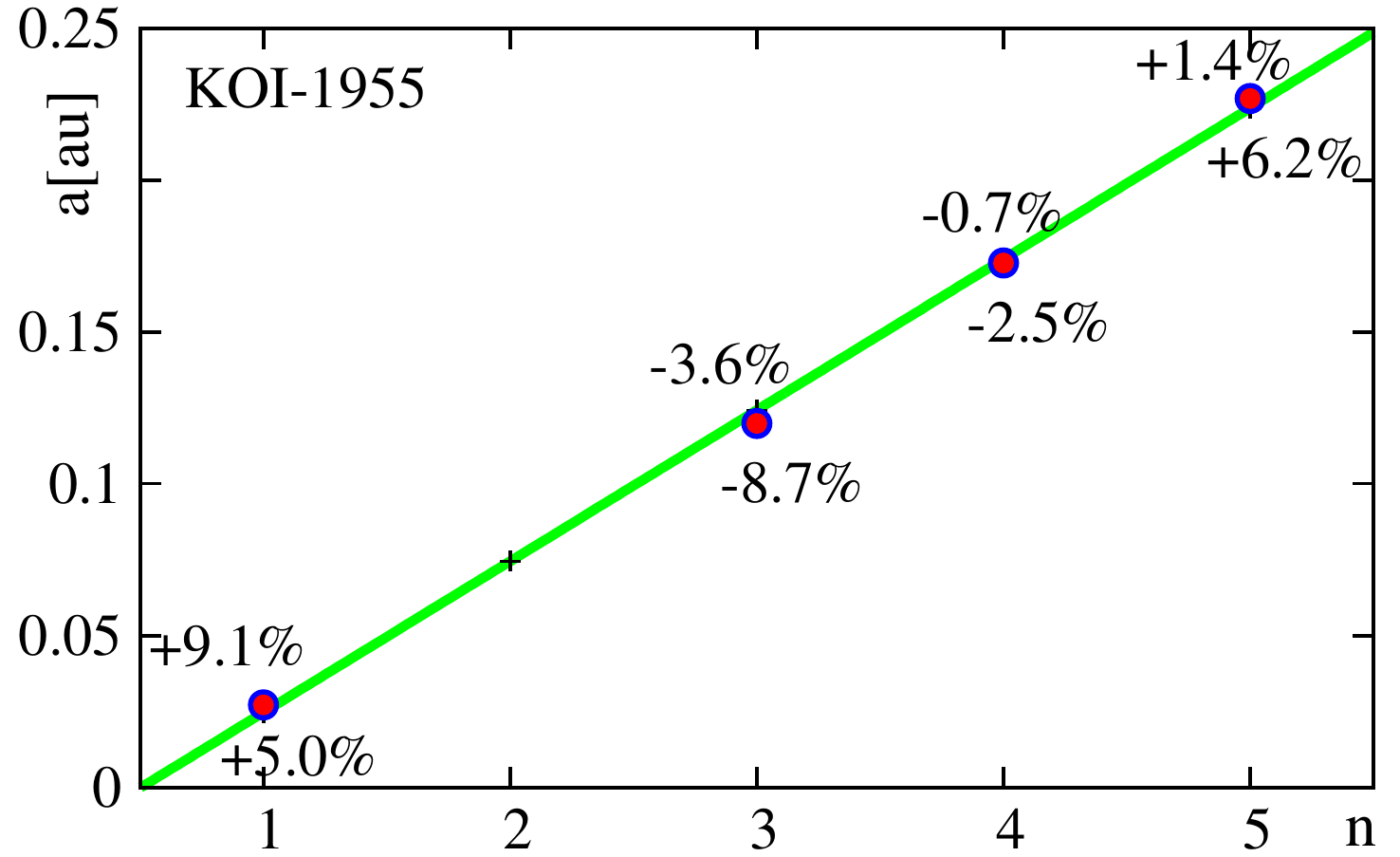}
}
}
\vbox{
\centerline{
\includegraphics[width=0.310\textwidth]{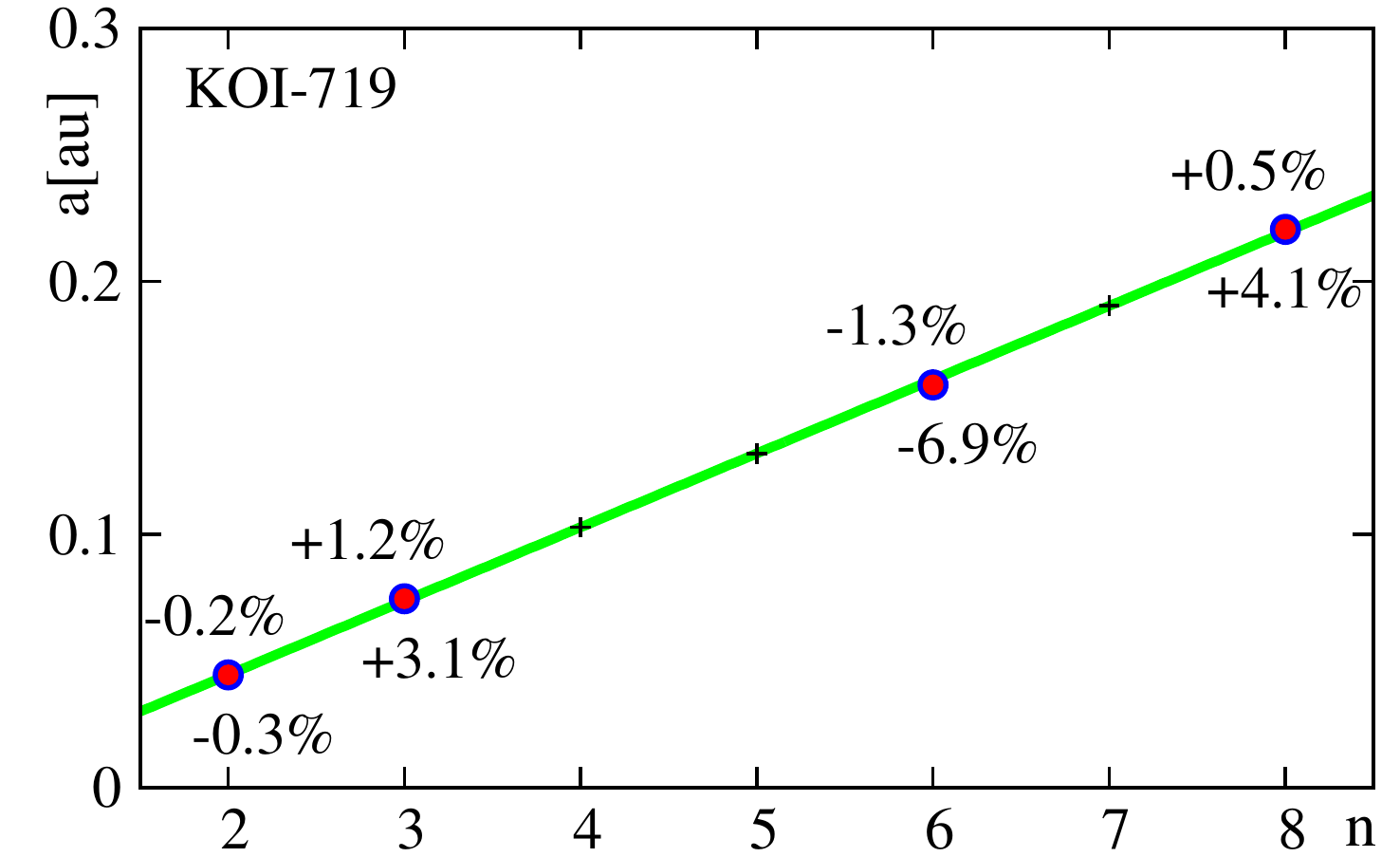}
\includegraphics[width=0.310\textwidth]{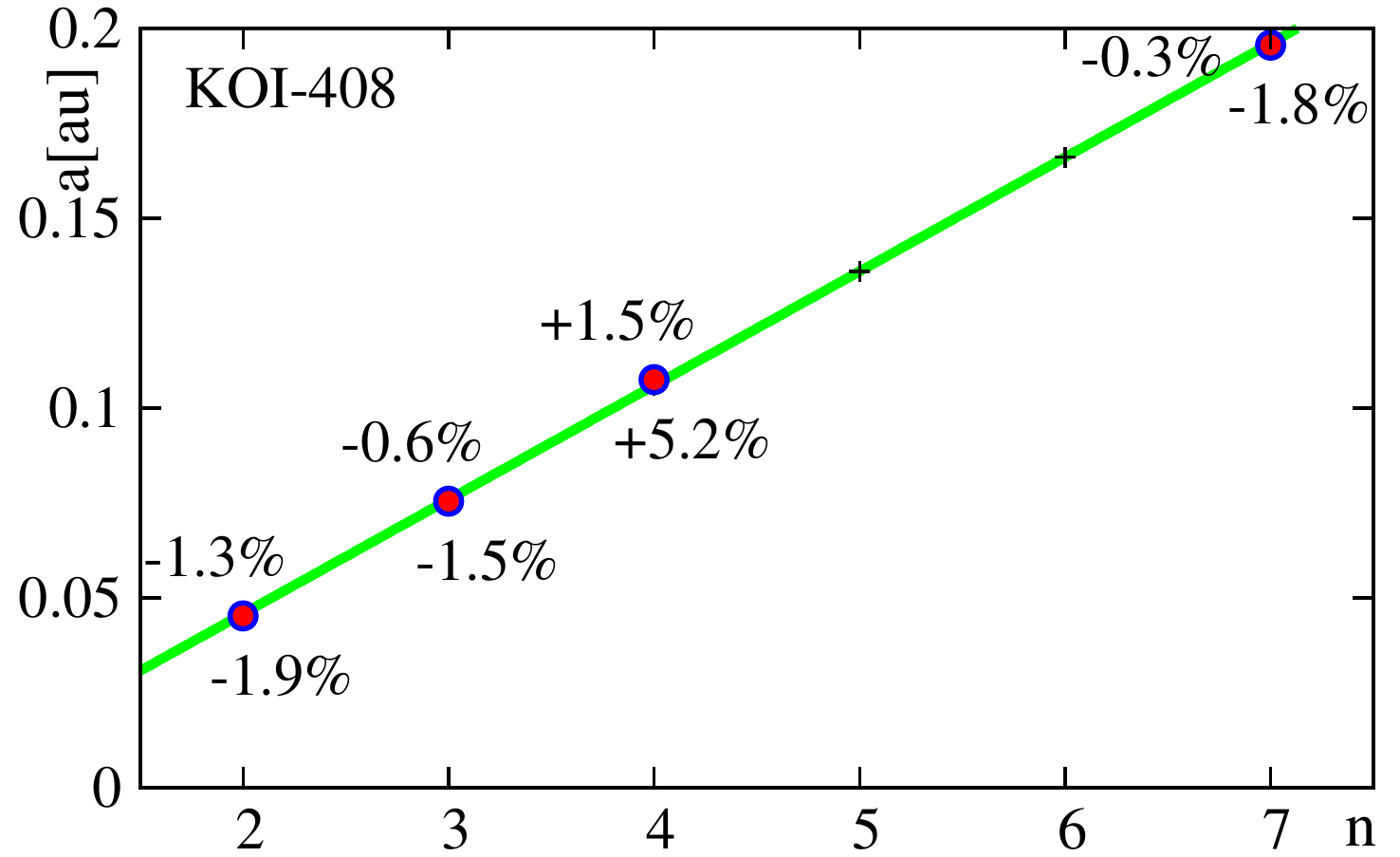}
\includegraphics[width=0.310\textwidth]{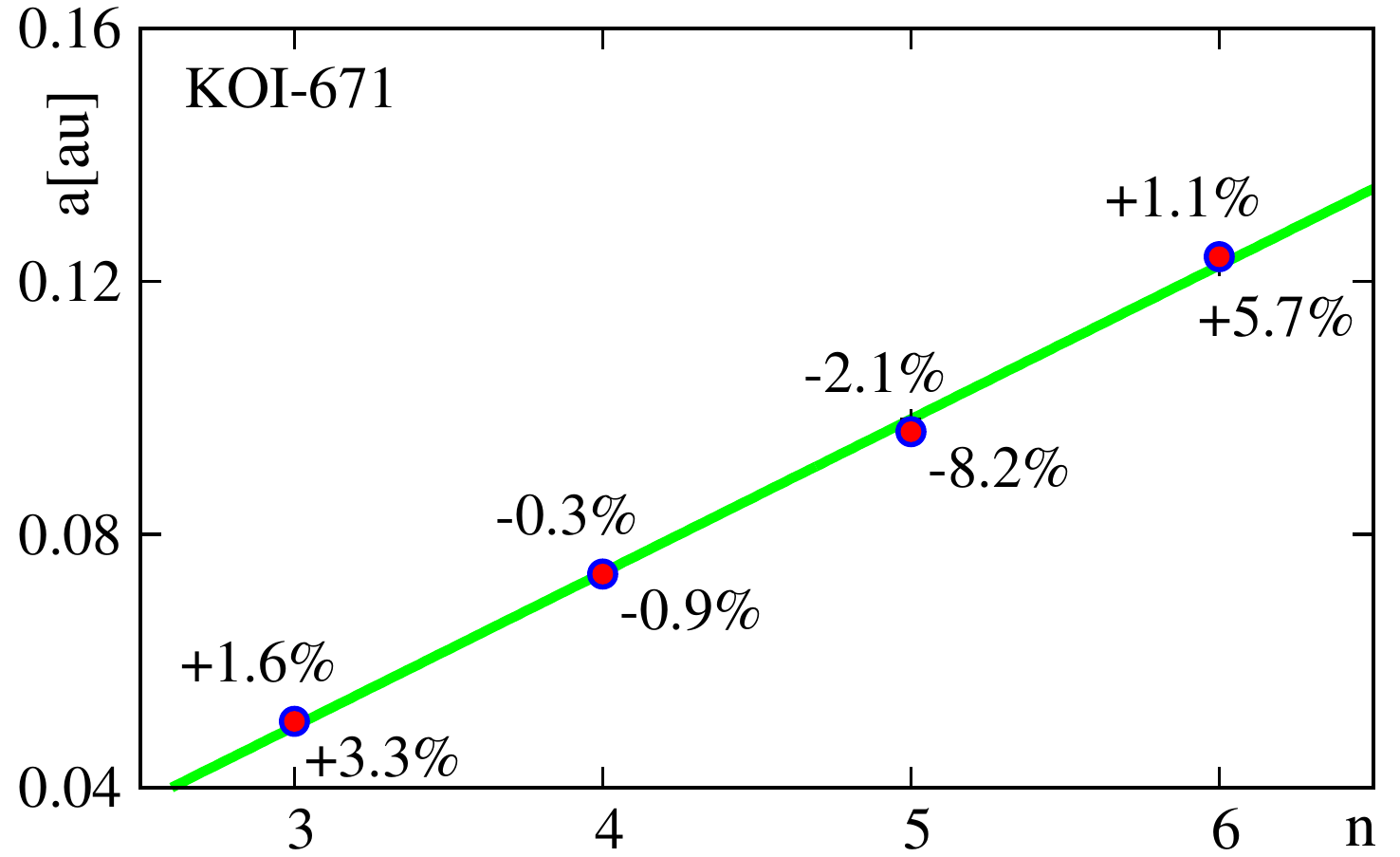}
}
}
\vbox{
\centerline{
\includegraphics[width=0.310\textwidth]{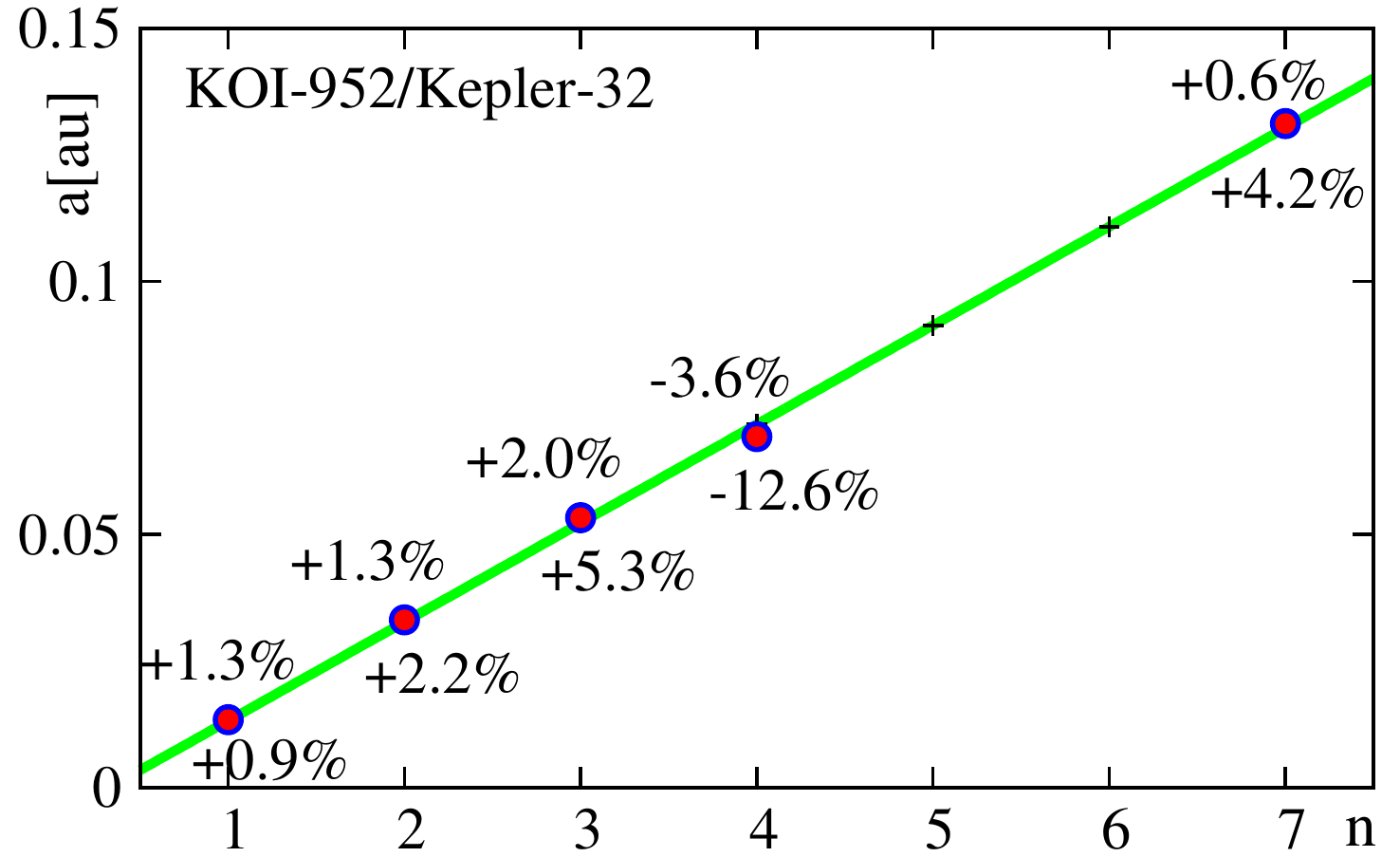}
\includegraphics[width=0.310\textwidth]{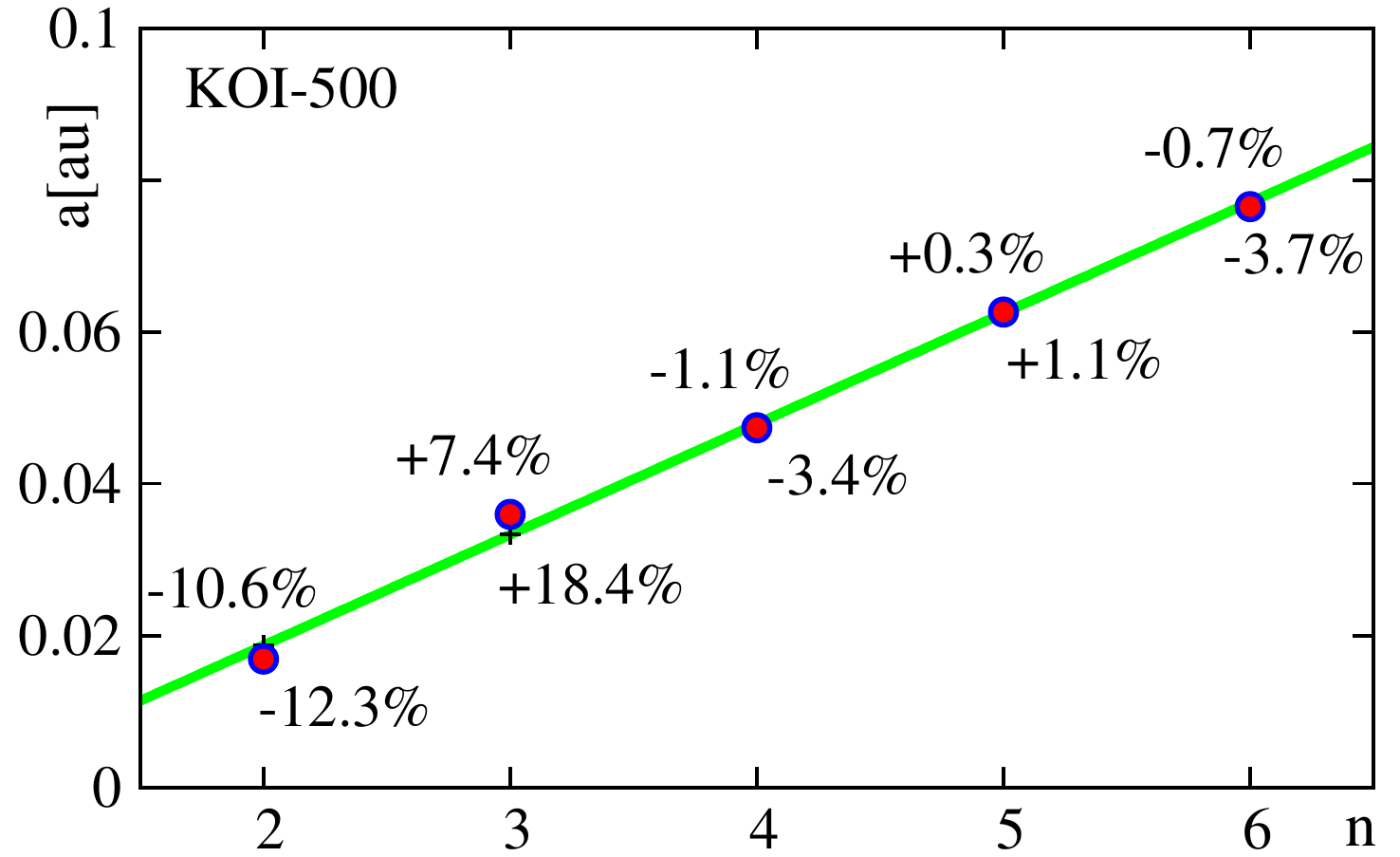}
\includegraphics[width=0.310\textwidth]{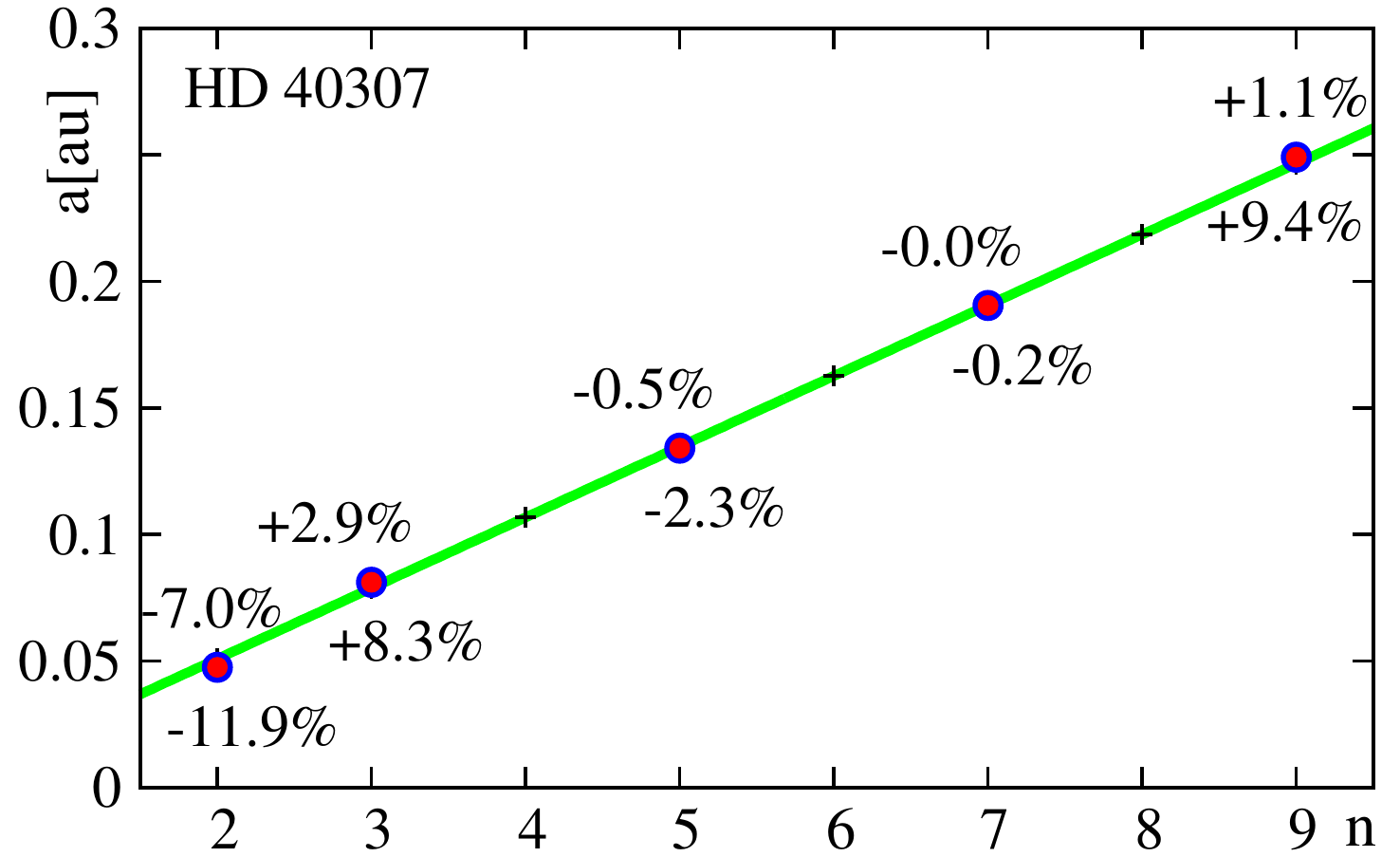}
}
}
\caption{
The $(n, a_n)$-diagrams of the best-fit linear solutions 
computed for chosen planetary systems. See the
text for details.
}
\label{fig:chosen_an}
\end{figure*}

\begin{figure*}
\vbox{
\centerline{
\includegraphics[width=0.310\textwidth]{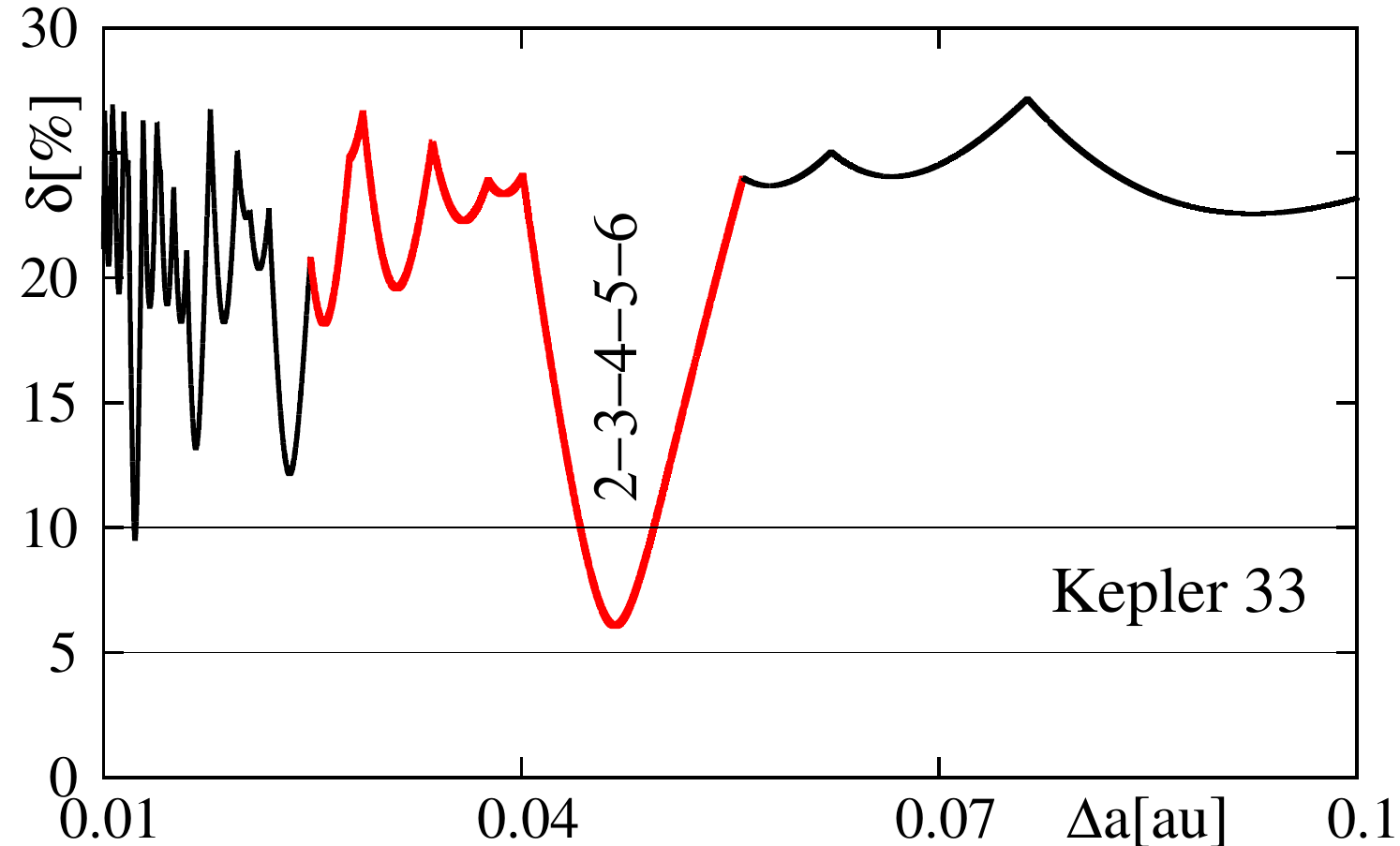}
\includegraphics[width=0.310\textwidth]{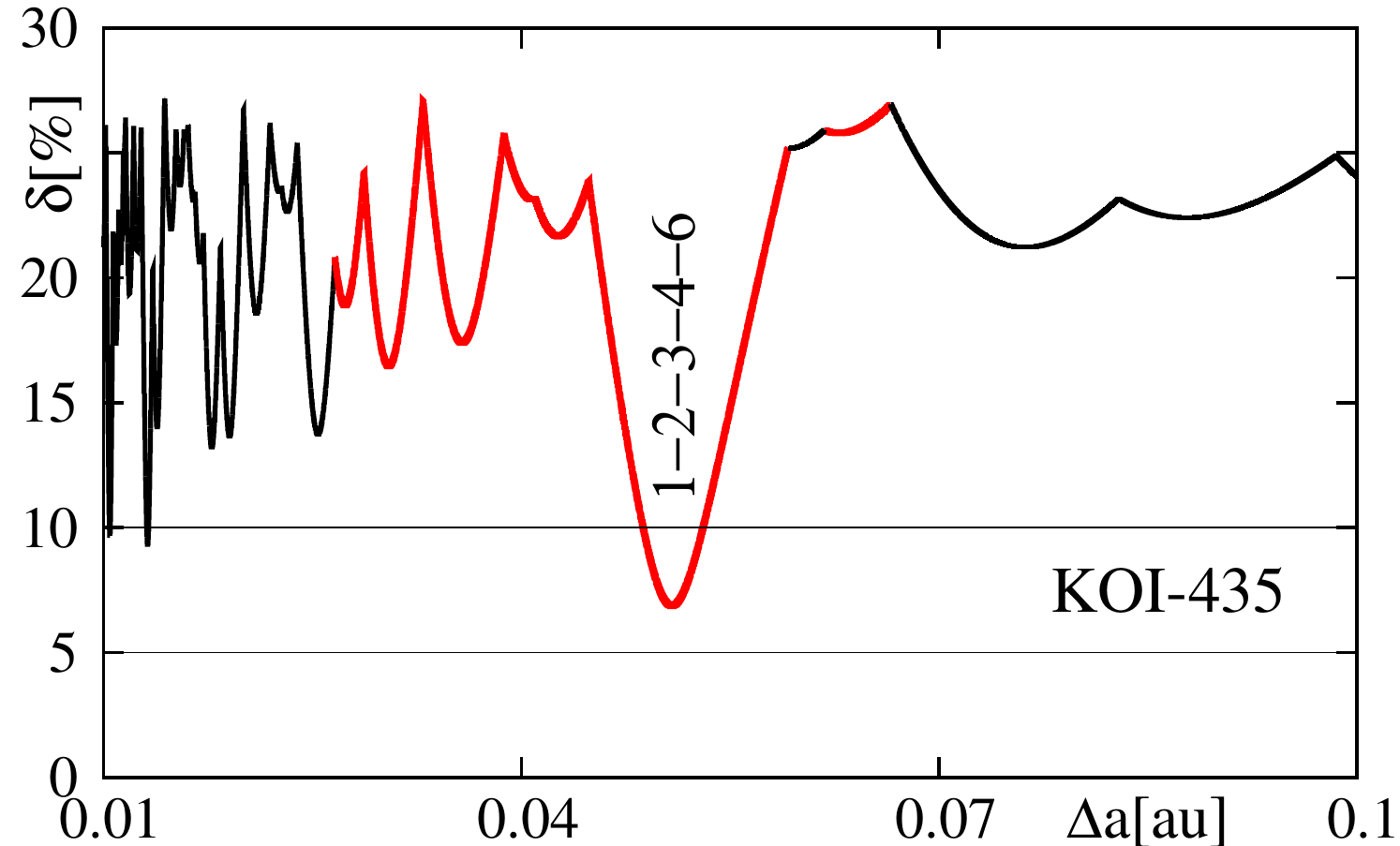}
\includegraphics[width=0.310\textwidth]{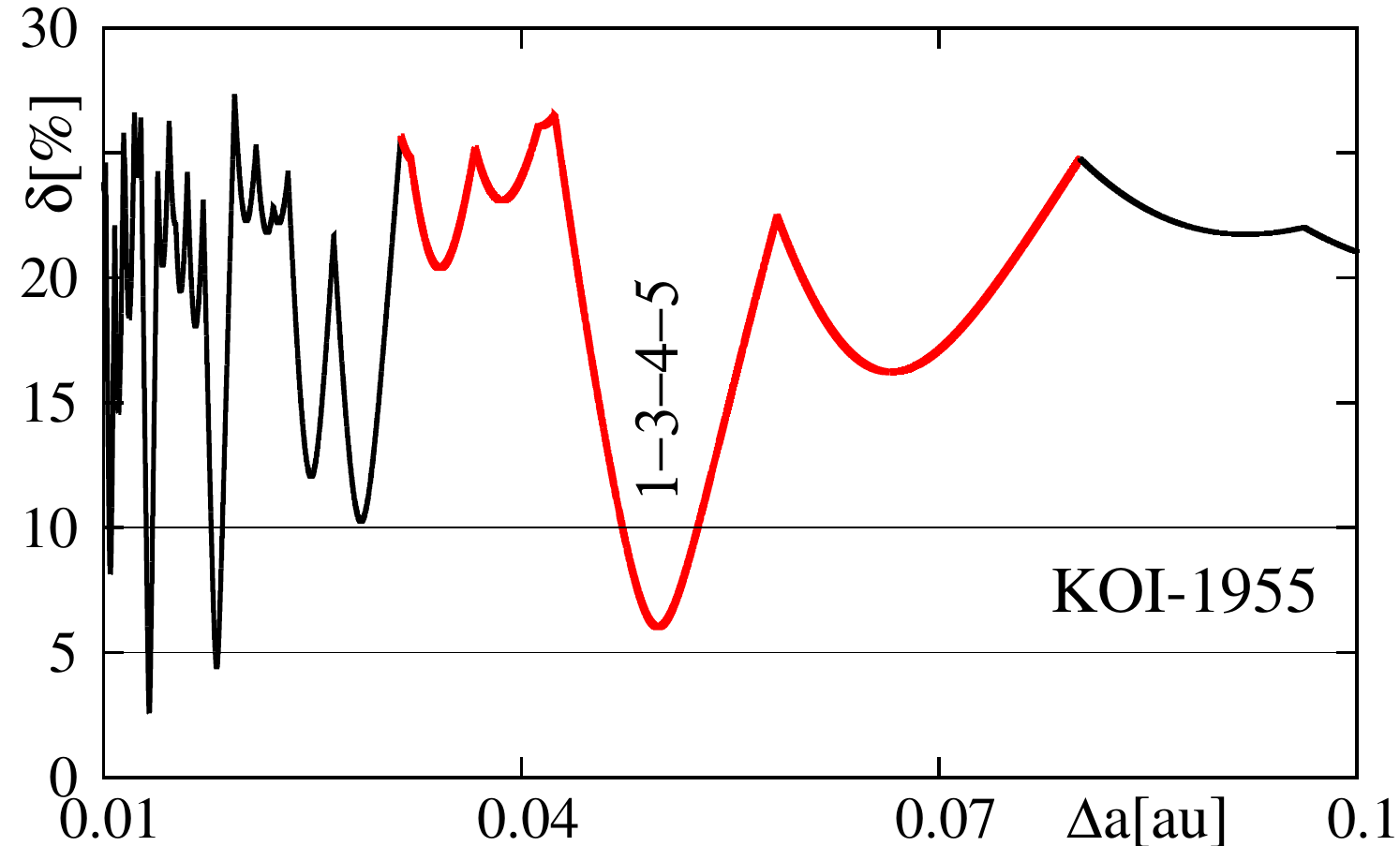}
}
}
\vbox{
\centerline{
\includegraphics[width=0.310\textwidth]{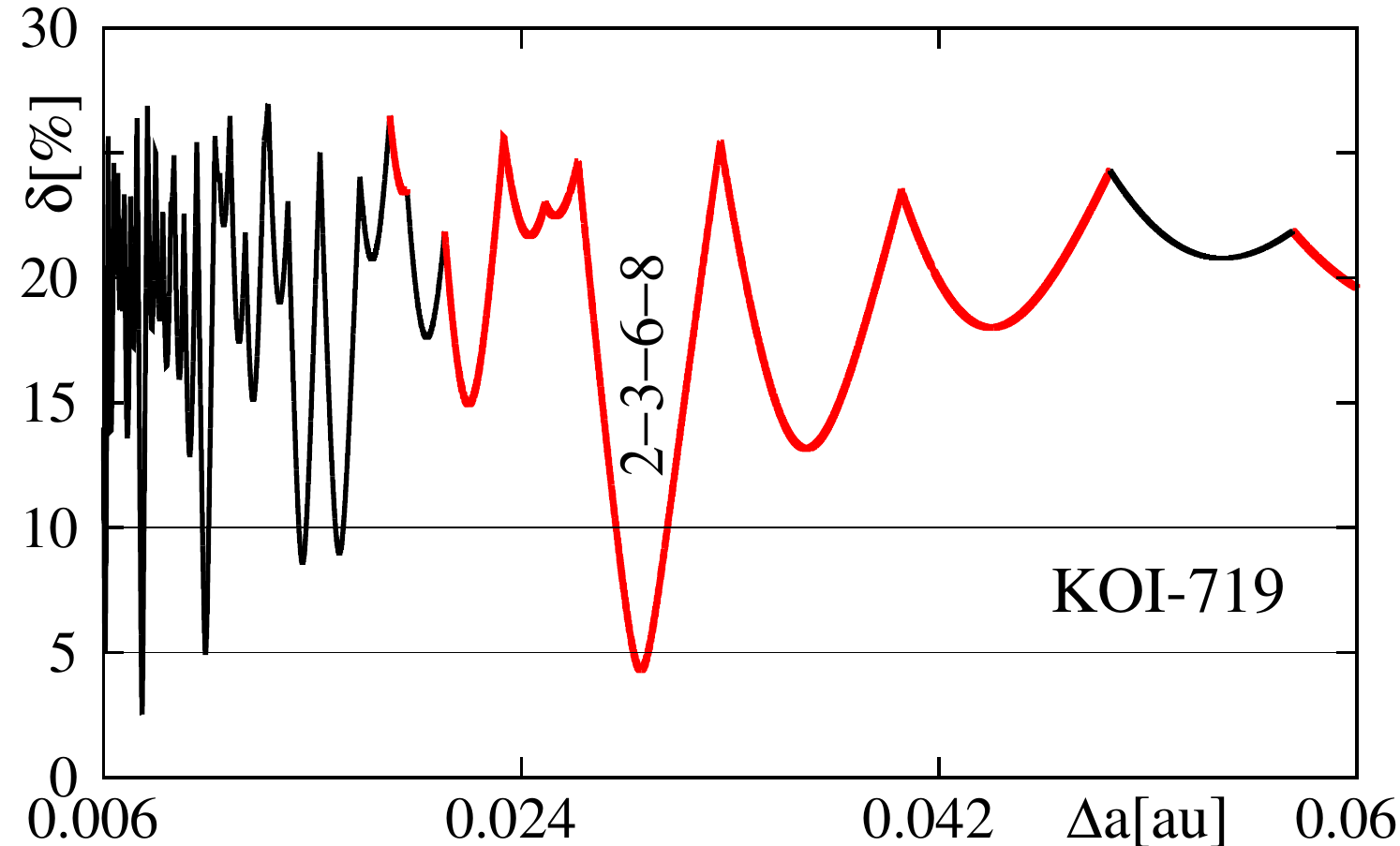}
\includegraphics[width=0.310\textwidth]{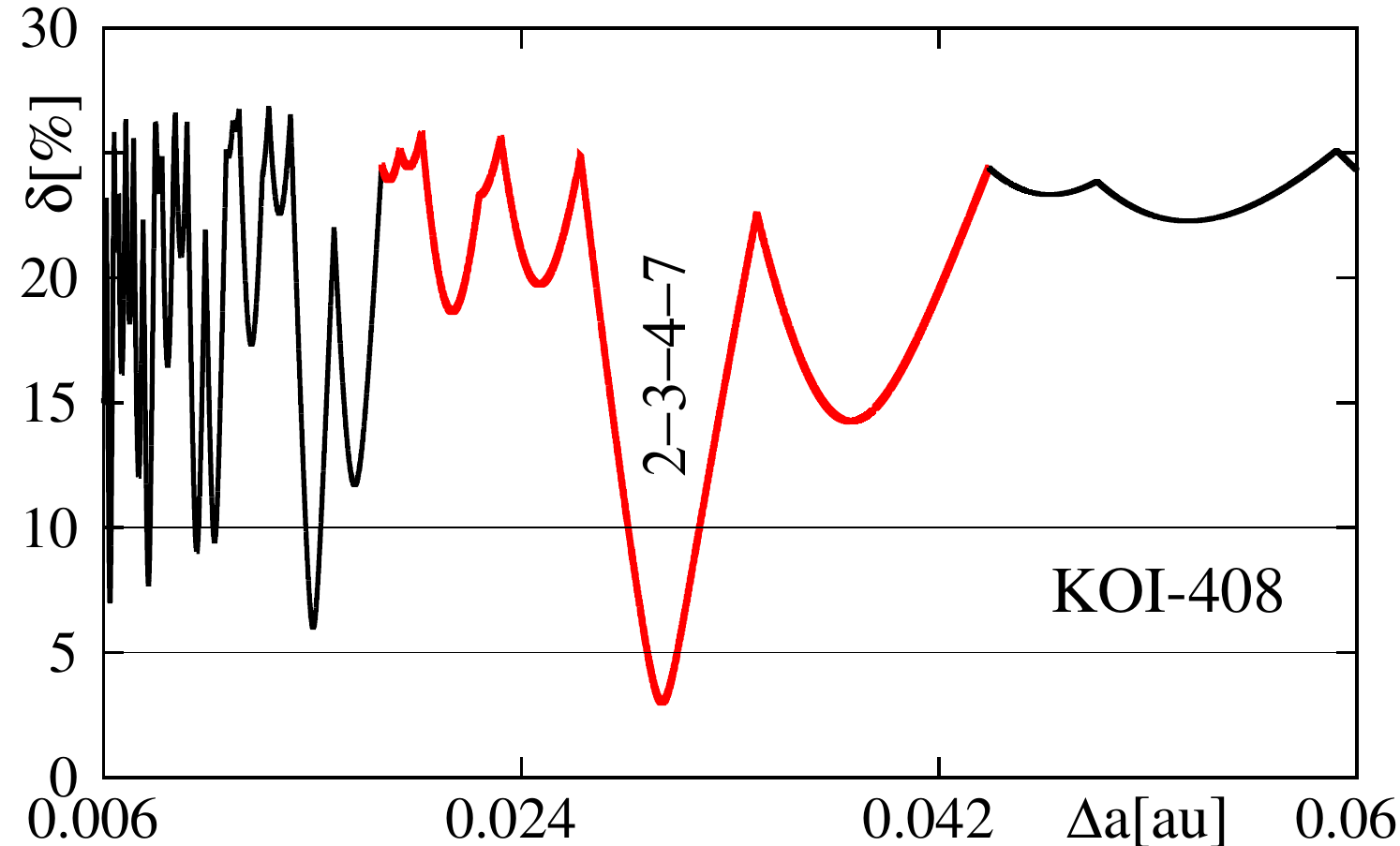}
\includegraphics[width=0.310\textwidth]{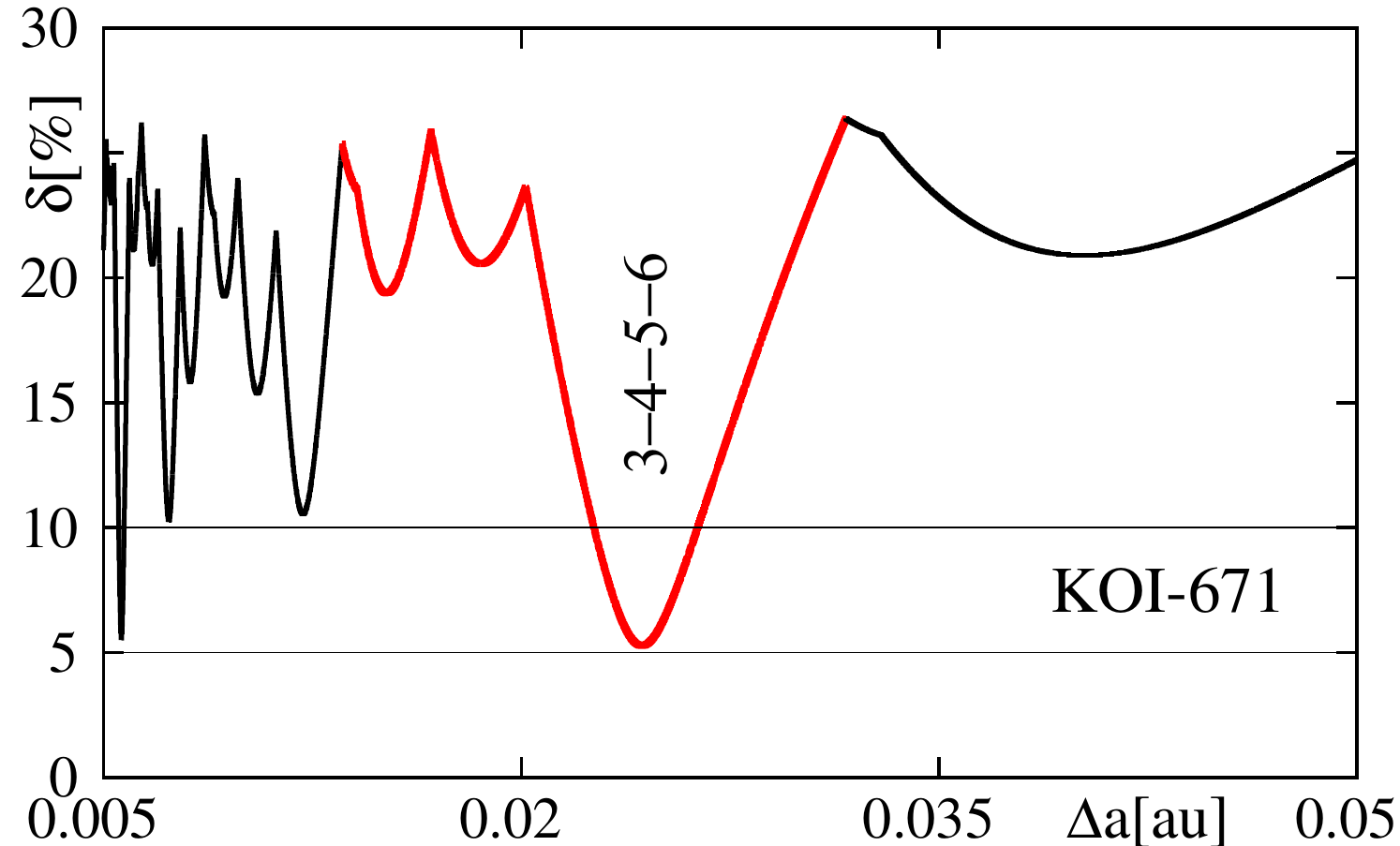}
}
}
\vbox{
\centerline{
\includegraphics[width=0.310\textwidth]{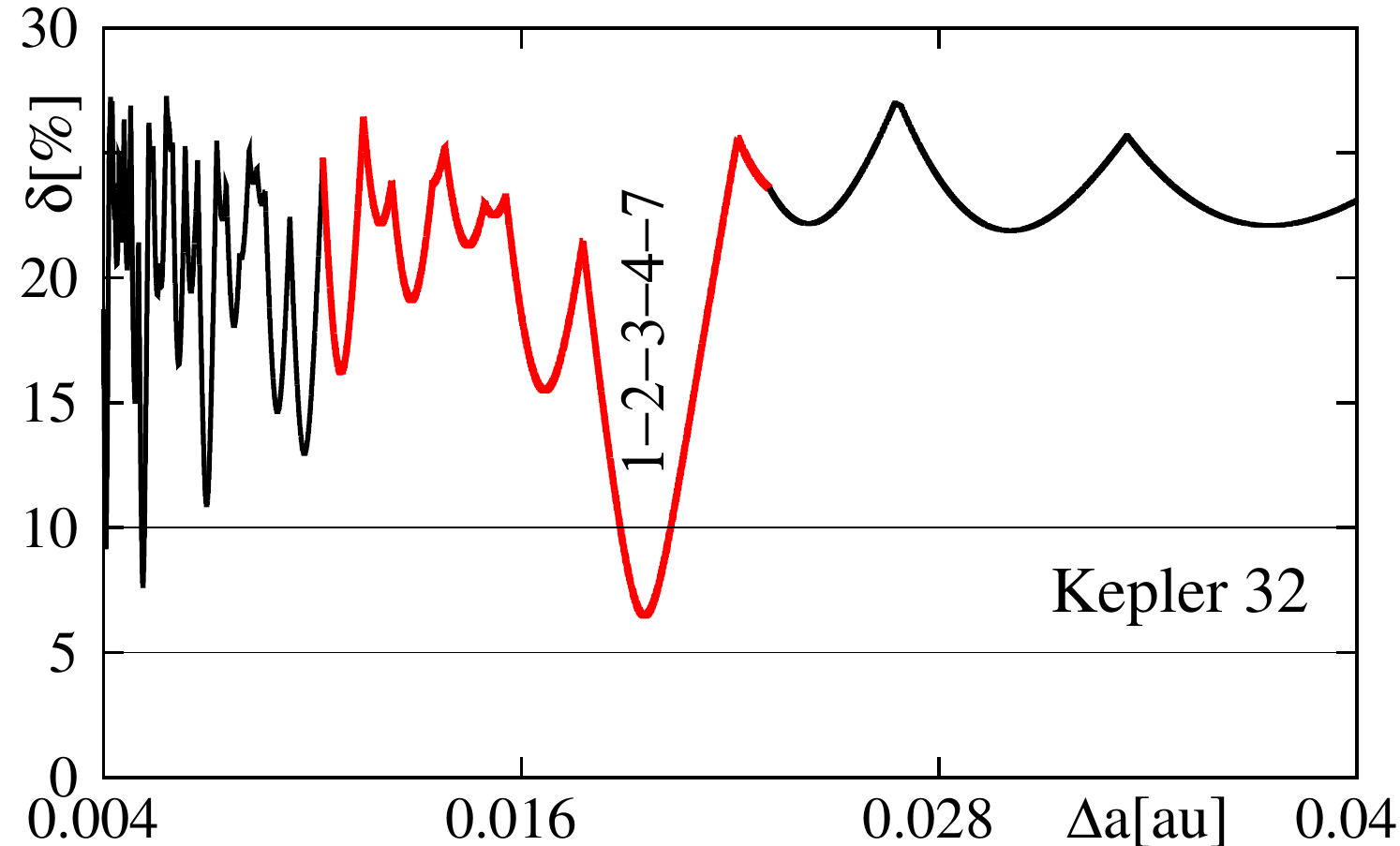}
\includegraphics[width=0.310\textwidth]{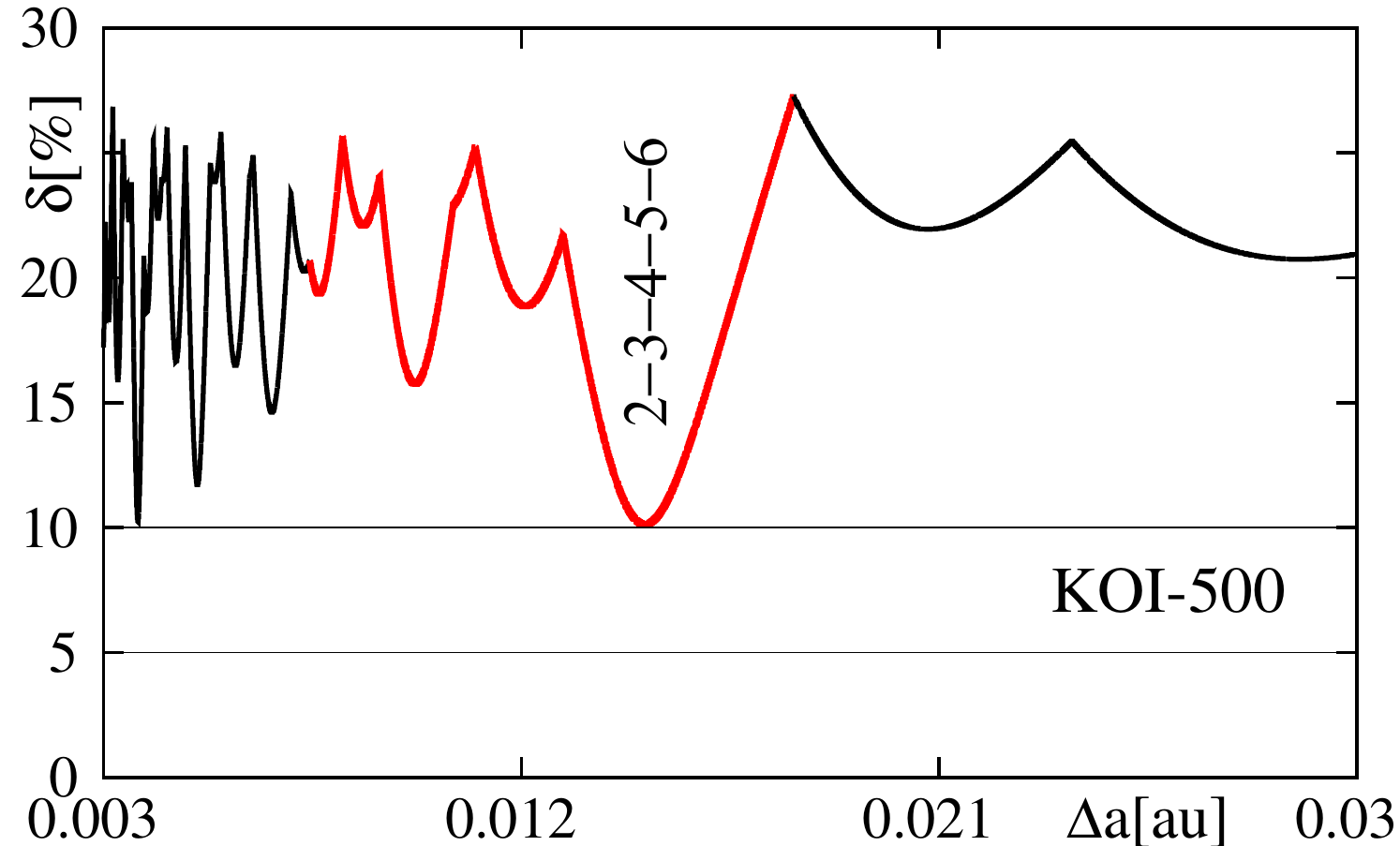}
\includegraphics[width=0.310\textwidth]{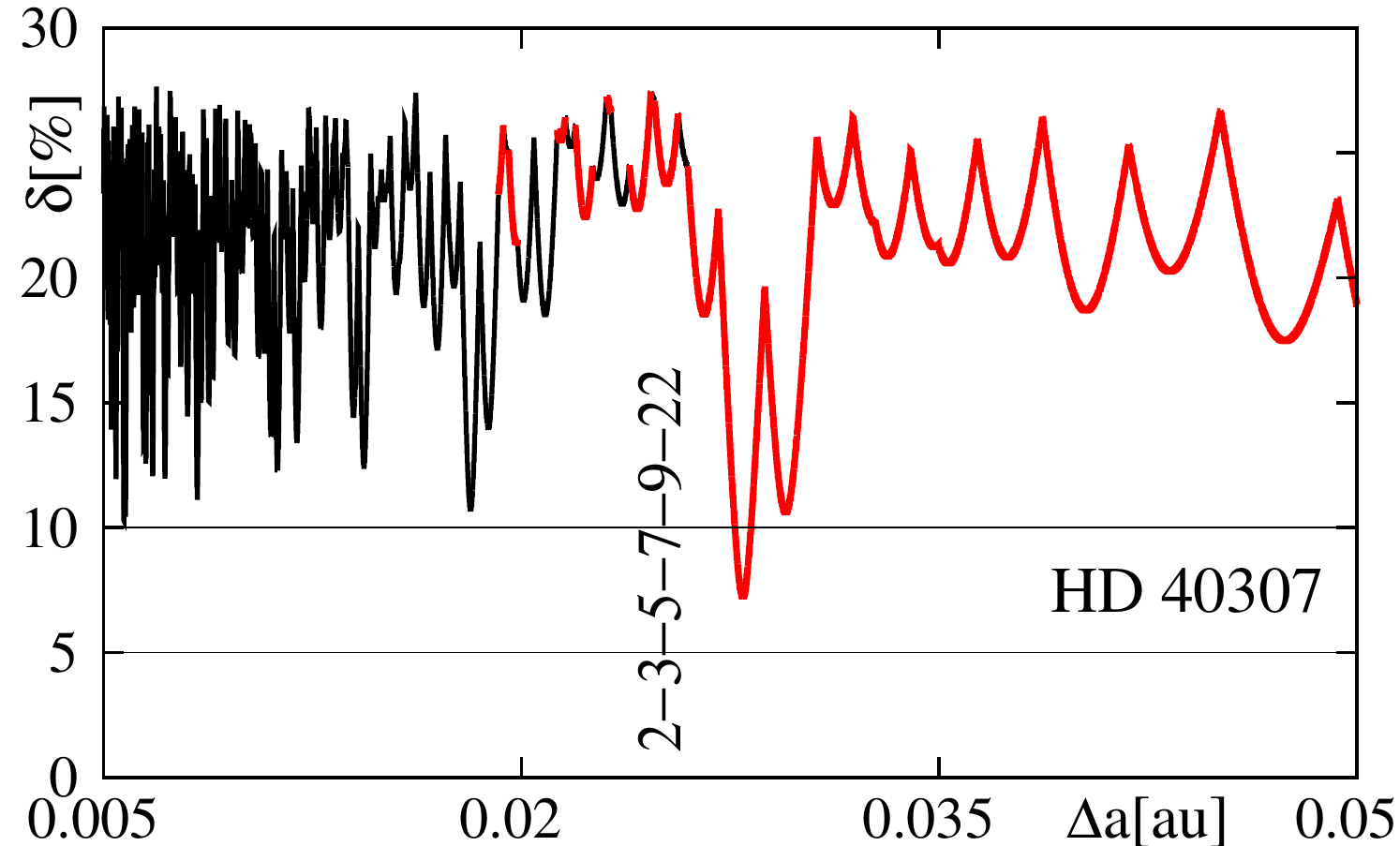}
}
}
\caption{
Goodness of the linear fit $\delta$ as a function of $\Delta\,a$.  Each
panel is for one system.  
}
\label{fig:chosen_delta2}
\end{figure*}

%
The sample consists of $20$ systems (including Kepler~33). The results are gathered in Table~\ref{tab:tab1}.  
Its columns
display the name of the star, its mass, the number of planets, the
reference, $\Delta\,a$, $a_1$, $\delta$, $f_{2/3}$, $f_1$ (False Alarm Probabilities, FAPs, defined below) and a sequence
of $n(i)$. 
A few planetary systems have more than one record. 
Figure~\ref{fig:chosen_an} shows the $(n, a_n)$-diagrams for $9$ chosen systems. 
For a reference, 1-dim scans of $\delta(\Delta\,a)$ are presented in
Fig.~\ref{fig:chosen_delta2}. The choice of systems to be shown in Figs.~\ref{fig:chosen_an} and~\ref{fig:chosen_delta2} was made on basis of a few criteria, which are: as low $\delta$ and FAP as
possible, and as few gaps as possible. Because this is a multiple-criteria choice it has to be, to some degree, arbitrary.

KOI-435 is a system with five planetary candidates in orbits of $a \lesssim 0.4\,\au$
and the sixth object, for which only one
transit was observed, is much more distant from the star.  Here, we take into account only five inner
candidates.  Fig.~\ref{fig:chosen_delta2} reveals that the linear
model corresponds to the minimum of $\delta$ around $\Delta\,a \approx
0.05$~au, which is close to the value for the Kepler-33 system.  The
quality of this model is very good, $\delta \approx 7\,\%$.  Indexes of
the planets are $1-2-3-4-6$, hence there is a gap between planet~4 and
planet~6.  We did not \corr{find} any better nor alternative solution.
It is not yet possible to say if such a gap should be filled by yet undetected planet.
The question is if such gaps are frequent outcomes of physical processes leading to discussed architecture. If they are rare one might expect a planet with $n=5$ in the KOI-435, otherwise we cannot make any predictions.
Figure~\ref{fig:chosen_an} shows $(n, a_n)$-diagram for this 
sequence.  This system seems very similar to the Kepler-33 system.  A
difference of $a_1$ means that the orbits of planets in KOI-435 are slightly
shifted, when compared to the Kepler-33 orbits.  In both cases, stellar
masses are known with $\sim 10\,\%$ uncertainty, which propagates into $\sim
3\,\%$ uncertainty of $a_n$, as well as of $\Delta\,a$ and $a_1$.

We estimate, that the remaining $7$ systems shown in Fig.~\ref{fig:chosen_delta2} obey the
linear law similarly well. The most interesting example here is KOI-500 with five
planets, which form a sequence $2-3-4-5-6$ (the same as Kepler~33).  All
planets reside within the distance of $0.08\,\au$ from the parent star.

The Kepler-31 system (not shown in Fig.~\ref{fig:chosen_delta2}) of four candidate planets, exhibits non-unique solutions ($\Delta\,a \approx 0.052\,\au$ and $\Delta\,a \approx 0.081\,\au$).
For both of them $\delta \approx 8\,\%$.  
Indexes of these models are $2-3-5-8$ and $2-3-4-6$,
respectively.
The next system, Kepler-11 has six planets. Two of its inner orbits are
separated by only $\sim 0.015\,\au$.  Other orbits are separated by $\sim
0.05\,\au$ except of the last one, which is relatively distant (separated by $\sim 0.2\,\au$ from the preceding planet). 
There are two possible solutions: 
$\Delta\,a \approx 0.0154\,\au$ ($\delta
\gtrsim 13\,\%$) and $\Delta\,a
\approx 0.052\,\au$ ($\delta \approx 12\,\%$).  
For the second case the indexes are $2-2-3-4-5-9$, i.e., two innermost planets have the same number $2$. The parameters
are almost the same as for KOI-435 and Kepler-31 (the first solution). The remaining members
of the group of systems not shown in Fig.~\ref{fig:chosen_delta2} exhibit relatively large $\delta$ or the best-fit models
have usually many "gaps".  Moreover,
in some cases, more than one model is possible (see
Tab.~\ref{tab:tab1}).

Having in mind systems with many gaps and/or large values of $\delta$, one might ask whether the linear ordering
might be just a matter of blind coincidence, like the widely criticised Titius-Bode
(TB)
rule. To check the linear rule on statistical grounds, we applied
the Monte-Carlo approach of \cite{Lynch2003}. He expressed the TB
model in the logarithmic scale, which can be directly used in our case.
We then analyse a random sample of $10^7$ synthetic orbits of
$
 a_n = a_1 + [ (n-1) + k\,y_n ] \Delta a,
$
where $y_n \in [-0.5, +0.5]$ is chosen randomly, while $k>0$ is a scaling parameter.
We optimize each synthetic system and compute the percentage of systems
for which $\delta<\delta_0$, where $\delta_0$ is for the observed
system. The resulting FAPs for $k=2/3$ and $k=1$ are displayed
as $f_{2/3}$, and $f_{1}$ in Table~1, respectively. We conclude that the random occurrence of the linear ordering in unlikely ($f_{2/3} \lesssim 10\,\%$, $f_1 \lesssim 5\,\%$) for approximately half of the sample. Nevertheless, these results are not definite, as the FAPs might depend on the sampling strategy \citep{Lynch2003}.

\corr{We would like to stress here that the linear ordering of orbits is not expected to be a universal rule which all systems would obey. We found that some of them are ordered according to this rule while some other systems from the sample are built differently. Our next step is to explain this}.

%
%
\section{Is the linear rule reflecting MMRs?} 
%
%
Since the linear spacing of orbits cannot be pure coincidence for all systems, there should be
a physical mechanism leading to this particular ordering of them. 
Searching for possible explanations of this phenomenon, we found that it may
appear naturally due to the inward, convergent migration of the planets interacting with
the remnant protoplanetary disk. The migration of two planets in a
gaseous disk has been studied in many papers \citep[e.g.,][]{Papaloizou2006,Szuszkiewicz2012}. It is known that the migration usually leads
to trapping orbits into the mean motion resonances (MMRs). It is reasonable
to foresee that systems with more planets might be trapped into chains of
MMRs, see Conclusions.
We ask now if there is any
combination of MMRs between subsequent pairs of planets resulting in the
linear spacing of the orbits.

There are no MMRs leading to {\em the exact linear
spacing} of the orbits ($\delta = 0$).  However, we can pick up easily
many different linear model possessing $\delta \sim 1\,\%$.  We examined synthetic planetary systems of $5$ and $6$ planets
involved in multiple MMRs.  We searched for such combinations of MMRs
which lead to the linear distribution of semi-major axes with no "gaps", like
in the Kepler~33 case.  We found many models with $\delta < 4\,\%$.  Let us
quote some interesting examples.  For a five-planet system, subsequent MMRs
$7:3$, $5:3$, $3:2$ and $4:3$ corresponds to a sequence of indexes
$2-3-4-5-6$ and $\delta \approx 2.4\,\%$.
Actually, this is very similar to the Kepler~33 system. Its planets are
close to the same resonances.

A proximity of a particular pair of planets $i$ and $i+1$ to a given 
MMR $q:p$, i.e., $P_{i+1}/P_i \approx q/p$ (where $q,p$
are relatively prime natural numbers), can be expressed
through
$
\epsilon_{i, i+1}(q, p) \equiv \left( {q \, P_i}/{p \, P_{i+1}} - 1 \right) \times 100\,\%.
$
For Kepler~33 one finds
$
\epsilon_{\idm{b}, \idm{c}} (7, 3) \approx 0.4\,\%,  
\epsilon_{\idm{c}, \idm{d}} (5, 3) \approx 0.8\,\%, 
\epsilon_{\idm{d}, \idm{e}} (3, 2) \approx 2.7\,\%,  
\epsilon_{\idm{e}, \idm{f}} (4, 3) \approx 3.2\,\%,
$
where the subsequent planets are called as b,~c,~d,~e and f, respectively.
Periods ratios of the first two pairs of planets are almost exactly equal to
rational numbers $7/3$ and $5/3$. For two more distant pairs, deviations
from $3/2$ and $4/3$ are slightly larger, still as small as $\sim 3\,\%$. 

If, in accord with the linear law, there existed one more innermost planet,
it would be involved in $7:1$ MMR with planet~b.  In such a case, the
six-planet sequence would correspond to the MMRs chain of $7:1$, $7:3$,
$5:3$, $3:2$, $4:3$ and $\delta \approx 2.2\%$.  Yet other MMRs between
planets $1$ and $b$ are possible ($6:1$, $8:1$, $9:1$, $11:2$), leading
to $\delta < 3\,\%$.  One more example of six planets involved in low
order MMRs are: $5:1$, $2:1$, $5:3$, $3:2$, $4:3$ with 
$\delta \approx
3.4\,\%$ (the first MMR could be also $6:1$, $9:2$); $7:1$, $5:2$, $5:3$, $3:2$, $4:3$ ($\delta \approx
3.4\,\%$); $9:2$, $9:4$, $5:3$, $3:2$, $4:3$ ($\delta \approx 3.6\,\%$);
$6:1$, $7:3$, $7:4$, $3:2$, $4:3$ ($\delta \approx 3.6\,\%$).  There
are many other solutions with higher order resonances and/or larger
$\delta$.  The most frequent MMRs in such sequences are $3:2$,
$4:3$, $5:2$, $5:3$, $2:1$ and $7:3$.
\corr{Considering the 4:3~MMR,
\cite{Rein2012} argue that it is difficult to construct 
this resonance on the grounds of the
common planet formation scenario. 
However, \cite{Rein2012} studied two-planet systems and their results might be not necessarily extrapolated for systems with more planets. Indeed, a recent paper by \cite{Cossou2013}
suggests quite opposite that forming the low-order MMRs,
and the 4:3 MMR in particular,
might be quite a natural and common outcome of a joint 
migration of planetary systems with low-mass members.}

\begin{table}
\centering{
\caption{
List of MMRs ($\epsilon < 2\,\%$). See the text for an explanation.
}
\label{tab:resonances}
\begin{tabular}{l c c c }
\hline
system & res. $(\epsilon [\%])$ & res. $(\epsilon [\%])$ & res. $(\epsilon [\%])$ \\
\hline
Kepler 33 & $7$b : $3$c $(+0.4)$ & $5$c : $3$d $(+0.8)$ & \\
KOI-435 & $5$b : $2$c $(-0.9)$ & $8$d : $5$e $(-0.5)$ & \\
KOI-1955 & $7$c : $4$d $(+1.2)$ & $3$d : $2$e $(-0.3)$ & \\
KOI-671 & $7$b : $4$c $(-0.9)$ & $3$c : $2$d $(+0.6)$ &  \\
Kepler 32 & $3$d : $2$e $(+1.2)$ & $13$e : $5$f $(-0.1)$ &  \\
KOI-500 & $3$c : $2$d $(-0.8)$ & $3$d : $2$e $(-1.3)$ & $4$e : $3$f $(-1.3)$ \\
HD 40307 & $9$b : $4$c $(+0.9)$ & $5$d : $3$e $(-1.7)$ & $3$e : $2$f $(+0.3)$ \\
KOI-730 & $4$b : $3$c $(-0.06)$ & $3$c : $2$d $(-0.1)$ & $4$d : $3$e $(-0.006)$ \\
KOI-94 & $14$b : $5$c $(+0.6)$ & $12$d : $5$e $(-1.3)$ & \\
Kepler 11 & $5$b : $4$c $(-1.1)$ & $7$c : $4$d $(+0.5)$ & $7$d : $5$e $(-0.8)$ \\
 & $5$f : $2$g $(-1.4)$ & & \\
Kepler 20 & $5$b : $3$c $(+1.0)$ & $9$c : $5$d $(+1.1)$ & $9$d : $5$e $(-0.2)$ \\
 & $4$e : $1$f $(+0.9)$ & & \\
KOI-510 & $11$b : $5$c $(+1.3)$ & $9$c : $4$d $(-1.8)$ & $12$d : $5$e $(-0.4)$ \\
KOI-623 & $3$c : $2$d $(-1.0)$ & $8$d : $5$e $(-0.5)$ &  \\
KOI-505 & $4$c : $3$d $(-1.1)$ & $5$d : $3$e $(+1.1)$ &  \\
Gliese 876 & $2$c : $1$d $(-1.6)$ & $2$d : $1$e $(-1.7)$ &  \\
HD 10180 & $5$c : $3$d $(-0.6)$ & $5$d : $3$e $(-1.6)$ & $3$e : $1$f $(-1.4)$ \\
 & $4$f : $3$g $(-1.8)$ & $9$g : $5$h $(-1.1)$ &  \\
\hline
\end{tabular}
}
\end{table}

Kepler~33 is not the only system whose planets are close to
MMRs.  In Table~\ref{tab:resonances} we gathered other systems
with at least two MMRs with $|\epsilon| < 2\,\%$.
For most systems from the studied sample, there are two
or even more resonant pairs. The KOI-730 system is a good example here, as all three pairs of planets exhibit almost exact
period commensurabilites \citep{Fabrycki2011b}: planets~b and c are close to $4:3$~MMR ($\epsilon
\approx -0.06\,\%$), planets c and~d lie in a vicinity of the $3:2$~MMR ($\epsilon
\approx -0.1\,\%$), and planets d and e are trapped in the $4:3$~MMR
($\epsilon \approx -0.006\,\%$).  Furthermore, planets~b and~d, as well as
planets~c and~f are very close to $2:1$~MMR, and planets~b and~e are close
to $4:1$~MMR.  This is an amazing example of a multiple, chain structure of
MMRs.  Still, it is not the only known system with all planets trapped
into multiple MMRs (i.e., with $|\epsilon| \approx 1\,\%$).  The Kepler~20 systems exhibits the following
chain of MMRs, $5:3$, $9:5$, $9:5$, $4:1$.  This implies that planets~b and
d are close to $3:1$~MMR. 

\corr{We do not attempt to study here whether a given system is involved in an exact MMR or only evolves close
to this MMR. The migration does not necessarily result in trapping super-Earth into exact MMRs. Indeed, there are several mechanisms proposed in the literature to explain systematic and significant deviations of orbits in multiple Kepler systems from the MMRs \citep[e.g.,][]{Rein2012b, Lithwick2012, Petrovich2013}.
}

\corr{An inward migration of already formed planets is not the only scenario, 
when an early history of a planetary system is considered. 
A migration of small ''pebbles'' may take place before 
they form a planet \citep{Boley2013,Chatterjee2013} 
or both migration and formation may occur simultaneously. 
Although this is a very complex issue, 
because different mechanisms have to be taken into account, 
trapping planets into MMRs seems to be a natural outcome 
of a dissipative evolution of a young planetary system.}

%
\section{Conclusions and discussions} 
%
Although in the sample of $20$ packed planetary systems there are stunning
examples of the linear architecture, not all studied systems could be
satisfactorily described by the proposed rule. One possible explanation is
that there exist additional planets in these systems, not yet detected
(due to unfavourable orbit orientation or too small radii) \corr{or the systems are trapped into such chains of MMRs, which do not 
necessarily imply the linear architecture. 
It is also possible that the migration 
was stopped before pairwise MMRs were attained by the orbits, for instance
due to relatively early disk depletion.}

Some of the systems exhibit multiple-resonant structure, which, as we
found here, might explain the linear spacing law.  This
means an occurrence of a chain of two-body MMRs. Remarkably, some of
combinations of MMRs imply indexing of the planets without gaps.   Nevertheless, there are many other combinations which may lead to sequences including ''gaps''. 

It is widely believed that a convergent migration of relatively small
planets within protoplanetary disk or due to tidal interaction with the outer disk leads to trapping the planets into
MMRs.  Still, the underlying astrophysics is very complex \citep[][]{Paardekooper2013, Quillen2013}.  
We performed preliminary numerical
studies of a simple model of planet-disk interaction \corr{\citep{Moore2013}}.  We found that multiple-resonance capture is very likely, indeed. 
Recently, \cite{Moore2013} showed that, for appropriately chosen initial semi-major axes and rates of migration, it is possible to simulate appearance of the chain of resonances in KOI-730. This result is encouraging for the explanation of the
linear spacing as the final outcome of relatively ''quiet'' and slow
migration of the whole, interacting systems towards the observed state. Obviously final chain of MMRs 
\corr{as well as $\Delta\,a$} depend on initial orbits as well as on disk
properties.  \corr{Longer migration at a given rate can result in smaller
$\Delta\,a$.}
Our finding might be also confirmed by the fact that in most Kepler systems
planets are not captured into {\em exact} MMRs --- but are found close to
them \citep[e.g.,][]{Jenkins2013}. 
Detailed studies of migration of multiple-planet systems are necessary
to tell which final states, determined by the observed architectures, are
likely.  We postpone this problem to future papers.

%
\section*{Acknowledgments}
%
We would like to thank Evgenya Shkolnik, Aviv Ofir, Guillem Anglada-Escud{\'e} and Stefan Dreizler for a discussion. \corr{We thank anonymous referee for remarks which helped us to improve the paper.}
This work was supported by the Polish Ministry of Science and Higher
Education, Grant N/N203/402739.  C.M.  is a recipient of the stipend of the
Foundation for Polish Science (programme START, editions 2010 and 2011). 
This research was supported by 
Project POWIEW,
co-financed by the European Regional Development Fund under the Innovative
Economy Operational Programme.
%
%
\bibliographystyle{mn2e}
\bibliography{ms}
\label{lastpage}
\end{document}